\documentclass[aps,prl,superscriptaddress,reprint,nobibnotes]{revtex4-1}
\usepackage{amsmath}
\usepackage{amssymb}
\usepackage{graphicx}
\usepackage{hyperref}
\usepackage{url}
\usepackage{color,xcolor}
\usepackage{epstopdf}

\begin{document}

\title{Ferroelectric Ferrimagnetic LiFe$_2$F$_6$: Charge Ordering Mediated Magnetoelectricity}
\author{Ling-Fang Lin}
\thanks{L.F.L and Q.R.X contributed equally to this work.}
\affiliation{School of Physics, Southeast University, Nanjing 211189, China}
\author{Qiao-Ru Xu}
\thanks{L.F.L and Q.R.X contributed equally to this work.}
\affiliation{Department of Physics, Indiana University, Bloomington, Indiana 47405, USA}
\author{Yang Zhang}
\author{Jun-Jie Zhang}
\author{Yan-Ping Liang}
\affiliation{School of Physics, Southeast University, Nanjing 211189, China}
\author{Shuai Dong}
\email{Corresponding author. Email: sdong@seu.edu.cn}
\affiliation{School of Physics, Southeast University, Nanjing 211189, China}
\date{\today}

\begin{abstract}
Trirutile-type LiFe$_2$F$_6$ is a charge-ordered material with Fe$^{2+}$/Fe$^{3+}$ configuration. Here its physical properties, including magnetism, electronic structure, phase transition, and charge ordering, are studied theoretically. On one hand, the charge ordering leads to improper ferroelectricity with a large polarization. On the other hand, its magnetic ground state can be tuned from the antiferromagnetic to ferrimagnetic by moderate compressive strain. Thus, LiFe$_2$F$_6$ can be a rare multiferroic with both large magnetization and polarization. Most importantly, since the charge ordering is the common ingredient for both ferroelectricity and magnetization, the net magnetization may be fully switched by flipping the polarization, rendering intrinsically strong magnetoelectric effect and desirable function.
\end{abstract}

\maketitle

\textit{Introduction.-}
Multiferroics, with spontaneous magnetic order and charge dipole order, are not only physically interesting as emergent systems but also potentially useful as functional materials \cite{Cheong:Nm,Dong:Ap,Fiebig:Nrm}. Despite the significant advances in the past decades, it remains challenge to obtain the desirable physical properties, i.e. the coexistence of large polarization ($\textbf{P}$), large magnetization ($\textbf{M}$), strong and sensitive magnetoelectric coupling, at room temperature ($T$) \cite{Fiebig:Nrm,Dong:Ap}.

Classified by the underlying physical mechanisms, there are several routes to access the multiferroicity. For example, the origins of magnetism and ferroelectricity in BiFeO$_3$ are independent of each other, i.e. both $\textbf{P}$ and antiferromagnetic $\textbf{L}$ are primary order parameters. Thus, the coupling between $\textbf{P}$ and $\textbf{L}$ has to be indirect, mediated via the third ingredient, usually the structural distortion. In contrast, in magnetic ferroelectrics (the so-called type-II multiferroics), e.g. TbMnO$_3$ \cite{Kimura:Nat,Dong:Mplb} and CuO \cite{Kimura:Nm,Giovannetti:Prl11,Jin:Prl}, the ferroelectricity directly originates from the spiral magnetism, i.e. $\textbf{L}$ is a primary order parameter but $\textbf{P}$ is not. Such inequivalent roles make it rather difficult to control magnetism via electric methods, although it is easy to control $\textbf{P}$ via magnetic field.

To overcome the drawbacks of these two main branches, other new routes were explored in recent years. For example, the magnetic geometric ferroelectrics, e.g. hexagonal $RM$O$_3$ \cite{Aken:Nm,Xu:Mplb,Pang:Nqm,Lin:Prb16}, Ca$_3M_2$O$_7$ \cite{Benedek:Prl}, and Ba$M$F$_4$ \cite{Ederer:Prb06,Zhou:Sr15} ($R$: trivalent rare earth or Y; $M$: magnetic transition metal like Mn or Fe), have drawn much attentions. In these materials, the ferroelectricity is generated by collaborative multiple nonpolar modes of lattice distortion, i.e. $\textbf{P}$ is not a primary order parameter (thus different from the BiFeO$_3$ case). However, the origin of magnetism in these systems remains independent of $\textbf{P}$, and the magnetoelectric coupling remains mediated via lattice distortion modes (similar to the BiFeO$_3$ case).

Besides these branches, there is one more possible route based on charge ordering (CO) to generate multiferroicity \cite{Brink:Jpcm,Yamauchi:Jpcm}. The first proposed material was Lu$_2$FeO$_4$, which was reported to be ferroelectric induced by Fe$^{2+}$/Fe$^{3+}$ ordering between layers \cite{Ikeda:Nat,Subramanian:am,Nagano:Prl,Xiang:Prl07,Zhang:Prl07}. But following works questioned its ferroelectricity \cite{Angst:rrl,Groot:Prl12,Niermann:Prl12}. Other proposed CO multiferroics include Fe$_3$O$_4$ \cite{Alexe:Am} and Pr$_{0.5}$Ca$_{0.5}$MnO$_3$ \cite{Efremov:Nm,Giovannetti:Prl}. However, due to their very narrow band gaps (in fact Fe$_3$O$_4$ is a half metal above the CO Verwey point $124$ K \cite{Fonin:Prb,Liu:Nqm}), the low resistivity and thus serious current leakage make the experimental evidences of their ferroelectricity not fully convincing.

Recently, a new mechanism for magnetoelectric coupling based on the carrier-mediated field effect was proposed, which can be abstractly expressed as $(\nabla\cdot\textbf{P})(\textbf{M}\cdot\textbf{L})$ \cite{Weng:Prl}. However, it is challenging in experiments to fabricate the [111]-orientated BiFeO$_3$ few-layers, as designed in Ref.~\cite{Weng:Prl}.

Besides multiferroic oxides, there are many possible hidden multiferroics in fluorides, which can provide a fertile field to find materials with desired properties \cite{Scott:Jpcm,Yamauchi:Prl,Ederer:Prb06,Zhou:Sr15}. In this work, we will predict a new bulk multiferroic, trirutile-type LiFe$_2$F$_6$ \cite{Xu:Bok}, which can realize the desired multiferroicity based on CO mechanism but avoid the fabrication of artificial heterostructures. Although it has been synthesized nearly half a century ago \cite{Portier:Cras} and investigated as an electrode material in lithium ion batteries recently \cite{Zheng:Elechem,Liao:Jes1,Liao:Jes2}, the possible multiferroicity of LiFe$_2$F$_6$ has not been touched yet.

\begin{figure}
\centering
\includegraphics[width=0.48\textwidth]{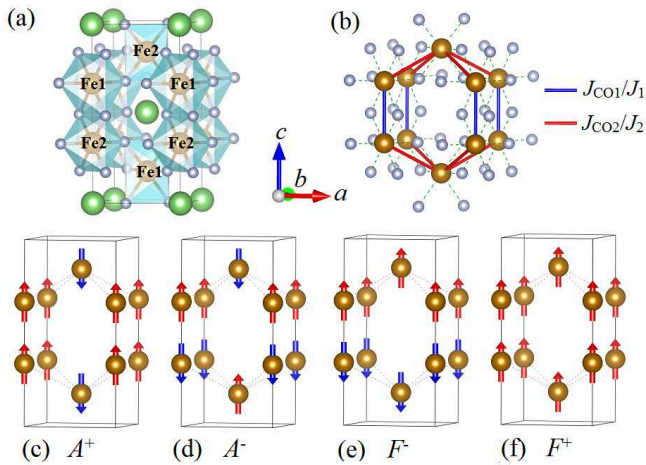}
\caption{(a) Crystalline structure of LiFe$_2$F$_6$. Brown: Fe; green: Li; silver: F. HSS: Fe1=Fe2. LSS: Fe1$\neq$Fe2. (b) The framework of Fe-F ions and the charge/magnetic exchange paths for $J_{\rm CO1}$/$J_1$ and $J_{\rm CO2}$/$J_2$. (c-f) Sketch of different magnetic orders of Fe spins. $F$ and $A$ ($+$ and $-$) stand for ferromagnetic and antiferromagnetic coupling between next-nearest neighbor (nearest neighbor) moments, respectively.}
\label{Fig1}
\end{figure}

\textit{Model system.-} LiFe$_2$F$_6$ forms the tetragonal crystal structure (see Fig.~\ref{Fig1}(a)). At high temperature, the high symmetric structure (HSS) without CO is $P4_2/mnm$ (No. 136). M\"{o}ssbauer spectrum measurement found the existence of Fe$^{2+}$ and Fe$^{3+}$ in LiFe$_2$F$_6$ above room-$T$ \cite{Greenwood:Jcsa,Fourquet:Jssc}, although the configuration of Fe$^{2+}$ and Fe$^{3+}$ could not be determined at that time. Later, Fourquet \textit{et al.} studied LiFe$_2$F$_6$ single crystal using X-ray diffraction, which revealed low symmetric structure (LSS, No. 102 $P4_2nm$) for the CO state \cite{Fourquet:Jssc}. Actually, the structural difference between HSS and LSS is quite subtle. In practice, the LSS can be also refined to HSS if the tolerance is increased a little bit.

The most relevant collinear orders of Fe spins include the so-called $A^+$, $A^-$, $F^-$, and $F^+$, as defined in Ref.~\cite{Shachar:Prb} [see Fig.~\ref{Fig1}(c-f)]. Neutron powder diffraction revealed the $A^+$ antiferromagnetism (Fig.~\ref{Fig1}(c)) below $105$ K \cite{Shachar:Prb,Wintenberger:Ssc}.

\textit{Method.-}
To reveal the multiferroicity of LiFe$_2$F$_6$, density functional theory (DFT) calculations are performed using the projector augmented wave (PAW) pseudopotentials as implemented in Vienna {\it ab initio} Simulation Package (VASP) code \cite{Kresse:Prb99,Blochl:Prb2}. To acquire accurate description of crystalline structure and electron correlation, the revised Perdew-Burke-Ernzerhof for solids (PBEsol) functional and the generalized gradient approximation plus $U$ (GGA+$U$) method are adopted \cite{Perdew:Prl08,Dudarev:Prb}. In addition, the hybrid functional calculations based on HSE06 exchange are used to check the band gap \cite{Heyd:Jcp,Heyd:Jcp04,Heyd:Jcp06}. The standard Berry phase method is employed to estimate the ferroelectric $\textbf{P}$ \cite{King-Smith:Prb,Resta:Rmp}. The nudged elastic band (NEB) method \cite{Henkelman:Jcp} is adopted to simulate the flipping of $\textbf{P}$ and estimate the upper-limit of energy barriers.

The cutoff of plane wave basis was fixed to $650$ eV, a quite high value due to the element Li. The Monkhorst-Pack $k$-point mesh is set to be $6\times6\times3$ for the minimal cell. Both the lattice constants and atomic positions are fully relaxed until the force on each atom is below $0.01$ eV/{\AA}. To account the influence of strain, the in-plane lattice constants are fixed while the length of $c$-axis is optimized, as well as atomic positions.

With energy coefficients extracted from DFT, model calculations are employed to simulate the phase transitions via Monte Carlo (MC) method. Although intrinsic magnetoelectricity exists in LiFe$_2$F$_6$ (to be studied later), their {\em phase transitions} are physical decoupled, which can be simulated independently. Our MC simulation can estimate the magnetic N\'eel temperature ($T_{\rm N}$) and CO transition temperature ($T_{\rm CO}$), which are important properties for multiferroics. The charge degree of freedom is mapped to a charge lattice model, characterized by a coefficient $C$ between $-1$ and $1$ (stands for Fe$^{(2.5+\frac{C}{2})+}$):
\begin{equation}
H_{\rm charge}=-J_{\rm CO1}\sum_{<i,j>}C_i{\cdot}C_j-J_{\rm CO2}\sum_{[i,k]}C_i{\cdot}C_k.
\end{equation}
The spin degree of freedom is mapped to a Heisenberg model, characterized by $\textbf{S}$:
\begin{equation}
H_{\rm spin}=-J_1\sum_{<i,j>}\textbf{S}_i\cdot\textbf{S}_j-J_2\sum_{[i,j]}\textbf{S}_i\cdot\textbf{S}_j-A\sum_i(S_i^e)^2,
\end{equation}
where $J_{\rm CO1}$/$J_1$ is the interaction between the nearest-neighbor ($<>$) charges/spins along the $c$-axis while $J_{\rm CO2}$/$J_2$ stand for the next-nearest-neighbor ($[]$) ones (see Fig.~\ref{Fig1}(b)). Considering the experimental fact that CO transition is above room-$T$ while magnetic transition is about $105$ K \cite{Greenwood:Jcsa,Fourquet:Jssc,Shachar:Prb}, here $J_{\rm CO1}$/$J_{\rm CO2}$/$C$ can be approximated to be spin-independent when studying the CO transition, and $J_1$/$J_2$/$\textbf{S}$ can all be fixed as constants ($|\textbf{S}|$ is normalized to $1$) when studying the magnetic transition. $S_i^e$ is the projection of spin to the magnetic easy (or hard) axis on $i$-site, and $A$ is magnetocrystalline anisotropy coefficient.

In MC simulation, periodic boundary conditions are used with lattice size $L\times L\times L$ ($L=8$). Larger sized lattices (e.g. $L$=$16$ or $32$) have also been tested to check the finite size effects, which are negligible. Specific heats ($C_{\rm v}$) and the charge/spin structure factor $C$(\textbf{k})/$S$(\textbf{k}) (Fourier transform of real space correlation function) \cite{Dong:Prb08,Zhang:Prm,Lin:FOP} are measured as a function of $T$ to character phase transitions. 

\textit{Results \& discussion.-}
The choice of $U_{\rm eff}$ for Fe's $3d$ orbitals may be important to obtain the correct physical properties in DFT calculation. Here, the value of $U_{\rm eff}$ is tested from $0$ to $6$ eV, as presented in details in Supplemental Materials \cite{Supp}. It is found that $U_{\rm eff}$=$4$ eV can lead to the best consistent with available experimental data, regarding the magnetism, CO, and lattice constants. This value also agrees with the previous empirical value used for Fe-based oxides with octahedra \cite{Weng:Prl,Zhang:Prb15}. Thus $U_{\rm eff}$=$4$ eV will be used in the following by default.

\begin{table*}
\centering
\caption{The optimized structural parameters ($a$/$c$ in the tetragonal notation), local magnetic moment ($M_1$ for Fe$1$ and $M_2$ for Fe$2$) within the default PAW sphere, Bader charge ($B_1$ for Fe$1$ and $B_2$ for Fe$2$), energy difference ($\Delta E$), band gap, net magnetization ($M$), and ferroelectric polarization ($P$) for various magnetic structures. The experimental values (Exp. for short) of lattice constants are also listed for comparison. The $A^+$ of LSS is taken as the reference for energy.}
\begin{tabular*}{0.96\textwidth}{@{\extracolsep{\fill}}lccccccc}
\hline
\hline
& $a$/$c$ ({\AA}) &  $M_1$/$M_2$ ($\mu_{\rm B}$) & $B_1$/$B_2$ (e) & $\Delta E$ (meV/Fe) & Gap (eV) &$M$ ($\mu_{\rm B}$/Fe)& $P$ ($\mu_{\rm C}$/cm$^2$)\\
\hline
$P4_2/mnm$ & & & & & & \\
\hline
$F^+$     & 4.674/9.091 & 4.13/4.13 & 12.19/12.19 & 83.05& 0.3 & 4.5 & --\\
$F^-$  & 4.657/9.283 & 4.21/4.20 &  12.19/12.19& 124.98& metallic & 0 & --\\
$A^+$ & 4.671/9.082& 4.09/4.10 &  12.19/12.19& 45.75& 0.4 & 0 & -- \\
$A^-$  & 4.658/9.263 & 4.19/4.12 & 12.20/12.20 &  87.70& metallic &  0  & --\\
Exp.~\cite{Portier:Cras} & 4.673/9.29  & --    & --    & --  & --  &   -- &   --\\
\hline
$P4_2nm$  & & & & & & \\
\hline
$F^+$    & 4.674/9.249 & 3.79/4.37 & 12.46/12.01&  40.98& 0.7 & 4.5 & 13.6 \\
$F^-$ & 4.669/9.304 & 3.78/4.38 & 12.47/12.01 & 40.98& 0.8  & 0 & 13.3\\
$A^+$& 4.666/9.253 & 3.25/4.35 &  12.46/12.02 & 0.00&  0.8 & 0 & 13.0 \\
$A^-$ & 4.666/9.291 & 3.72/4.35 & 12.47/12.02 & 0.50&  0.7 & 0.5 & 12.4\\
Exp.~\cite{Fourquet:Jssc} & 4.679/9.324  & --    & --    & --  &   -- &   -- &   -- \\
\hline
\hline
\end{tabular*}
\label{Table1}
\end{table*}

DFT calculated data for unstrained LiFe$_2$F$_6$ are summarized in Table~\ref{Table1}. Among all candidate states, the lowest energy one is always the $A^+$ (despite the structural choice), in agreement with the neutron study \cite{Shachar:Prb}. In addition, the calculated lattice constants for LSS with $A^+$ magnetism are quite closed to the experiment ones \cite{Portier:Cras,Fourquet:Jssc}. Furthermore, the CO state is obtained only for the LSS, evidenced by the disproportion of local magnetic moments and Bader charge (shown in Table~\ref{Table1}), while the HSS always gives the charge uniform (CU) result. For all calculated magnetic states, the energy of LSS is always lower than that of HSS. Therefore, our DFT calculation on LiFe$_2$F$_6$ gives fully consistent results comparing with available experimental data \cite{Portier:Cras,Fourquet:Jssc}, which provides a solid starting point for following calculations.

\begin{figure}
\centering
\includegraphics[width=0.45\textwidth]{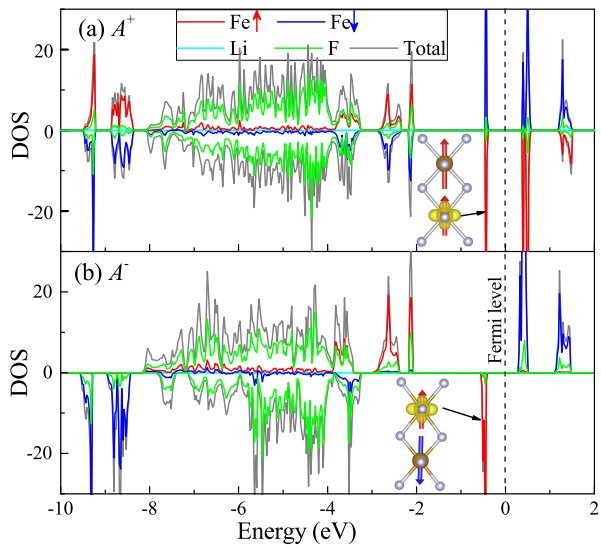}
\caption{DOS and atom-projected DOS (PDOS) of LiFe$_2$F$_6$ for the LSS. (a) $A^+$; (b) $A^-$. $\uparrow$/$\downarrow$ denote the spin directions. Insert: the electron cloud of the topmost valence band.}
\label{Fig2}
\end{figure}

For the ground $A^+$ state, LiFe$_2$F$_6$ is insulating, with a moderate band gap $0.8$ eV. Such a band gap has been further checked using the hybrid functional calculation based on HSE06 exchange \cite{Heyd:Jcp,Heyd:Jcp04,Heyd:Jcp06}, which leads to $1.0$ eV \cite{Supp}. Thus, comparing with other CO materials, LiFe$_2$F$_6$ is insulating enough to perform ferroelectric measurement. According to density of states (DOS) (Fig.~\ref{Fig2}), the bands near the Fermi level are mostly contributed by Fe's $3d$ orbitals. In particularly, the topmost valence band is contributed by Fe$^{2+}$'s upper Hubbard band of $d_{xy}$ orbital (if the $ab$-plane Fe-F bond is taken as the $z$-axis of F-octahedron), as visualized in the inserts of Fig.~\ref{Fig2}. The DOS for the first excited state $A^-$ is shown in Fig.~\ref{Fig2}(b), which owns a net $\textbf{M}$=$0.5$ $\mu_{\rm B}$/Fe.

The point groups of HSS and LSS are $4/mmm$ (nonpolar) and $4mm$ (polar), respectively. Thus, the CO transition is also the nonpolar/polar transition (i.e. ferroelectric Curie temperature $T_{\rm C}$=$T_{\rm CO}$). The dipole moment formed by Fe$^{2+}$-Fe$^{3+}$ pair can be estimated using the intuitive point charge model, which gives $12.4$ $\mu$C/cm$^2$. For comparison, the Berry phase method leads to $13.0$ $\mu$C/cm$^2$ along the $c$-axis for the $A^+$ state. Such an agreement implies the ideal CO driven ferroelectricity. For such an electronic ferroelectric state, the switching of $\textbf{P}$ can be realized by fast electron hopping between Fe$^{2+}$ and Fe$^{3+}$, instead of ion's slow displacements in most conventional ferroelectrics.

\begin{figure}
\centering
\includegraphics[width=0.48\textwidth]{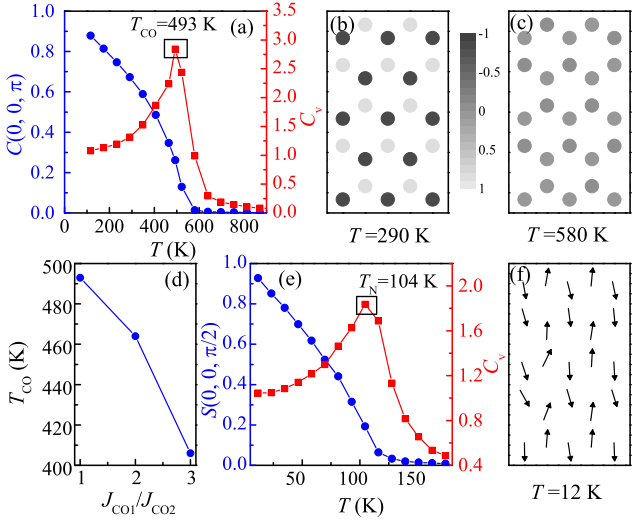}
\caption{MC results for models of LiFe$_2$F$_6$. (a) The charge structure factor $C(\textbf{k})$ and specific heat $C_{\rm v}$ as a function of $T$. (b-c) Contour plots for charge distribution (characterized by $C$) at room-$T$ and high-$T$. (d) $T_{\rm CO}$ as a function of $\frac{J_{\rm CO1}}{J_{\rm CO2}}$. (e) The spin structure factor $S(\textbf{k})$ and specific heat $C_{\rm v}$ as a function of $T$. (f) The real space spin pattern of MC snapshot obtained at low temperature. Here the vectors $\textbf{k}$ in $C(\textbf{k})$ and $S(\textbf{k})$ are reciprocal vectors of corresponding real space orders.}
\label{Fig3}
\end{figure}

By comparing the DFT energies of high and low symmetric $A^-$ (or $F^-$) states of CO/CU states, the combination $2J_{\rm CO2}+\frac{J_{\rm CO1}}{2}$ can be extracted for the charge lattice model. Note that it is impossible to stabilize any other CO pattern in the DFT self-consistent calculation. Thus, it is insufficient to obtain the individual values of $J_{\rm CO1}$ \& $J_{\rm CO2}$. Even though, with this coefficient $2J_{\rm CO2}+\frac{J_{\rm CO1}}{2}$, our MC simulation on lattice model can confirm that the ground state is indeed CO and the $T_{\rm CO}$ can be rather high. For example, $T_{\rm CO}$ is roughly estimated as $493$ K [Fig.~\ref{Fig3}(a)] when $J_{\rm CO2}=J_{\rm CO1}$. The contour plots of charge distribution at room-$T$ and high-$T$ can be seen in Fig.~\ref{Fig3}(b-c). $T_{\rm CO}$ deceases with increasing ratio of $\frac{J_{\rm CO1}}{J_{\rm CO2}}$ [Fig.~\ref{Fig3}(d)] but still higher than $400$ K when $\frac{J_{\rm CO1}}{J_{\rm CO2}}$=$3$, and the type of CO does not change with $\frac{J_{\rm CO1}}{J_{\rm CO2}}$.

By comparing the energies of different magnetic orders with an {\em identical structure}, the exchange interactions of the Heisenberg model are obtained as: $J_1=2.8$ meV and $J_2=-10.7$ meV \cite{Supp}. By incorporating the spin-orbit coupling (SOC) in the DFT calculation and comparing the energies of different spin axes \cite{Supp}, the magnetocrystalline easy axis is found to be along the $c$-axis, in agreement with the neutron study \cite{Shachar:Prb}, and the calculated $A=0.158$ meV/Fe. The MC-simulated $T_{\rm N}$ is about $104$ K (Fig.~\ref{Fig3}(e)), in good agreement with the experimental value ($105$ K) \cite{Shachar:Prb}. Such an agreement further confirms that $U_{\rm eff}$=$4$ eV is the best choice. Below $T_{\rm N}$, the real space spin pattern taken from MC snapshot, as shown Fig.~\ref{Fig3}(f), indicates the $A^+$ pattern.

However, the ground $A^+$ state is antiferromagnetic without a net spontaneous $\textbf{M}$. Instead, the first excited state $A^-$ is ferrimagnetic, leading to $0.5$ $\mu_{\rm B}$/Fe resulted from the Fe$^{2+}$($\downarrow$)-Fe$^{3+}$($\uparrow$) pair. Since the energy of $A^-$ state is only slightly higher ($0.50$ meV/Fe) than that of $A^+$, it is hopeful to tune the sign of superexchange $J_1$ and then the magnetic ground state.

Strain is a routinely useful method to tune physical properties of materials \cite{Schlom:Armr}. For example, strained EuTiO$_3$ becomes ferromagnetic and ferroelectric, although originally it is paraelectric and antiferromagnetic \cite{Fennie:Prl,Lee:Nat}.

\begin{figure}
\centering
\includegraphics[width=0.48\textwidth]{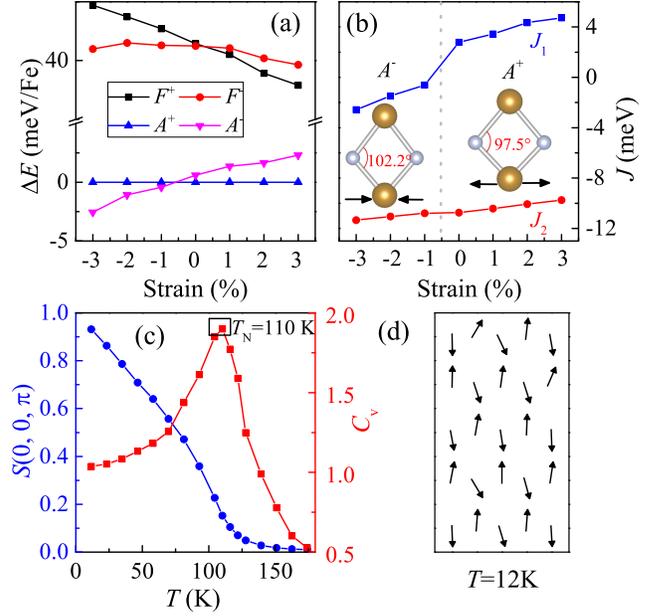}
\caption{Results for strained LiFe$_2$F$_6$. (a) Energies of the $F^+$, $F^-$, $A^+$, and $A^-$ states as a function of strain. The energy of $A^+$ state is the reference. (b) Strain dependent exchanges for the Heisenberg model. Insert: sketch of the bond angle of Fe$^{2+}$-F-Fe$^{3+}$ for the $J_1$ path. (c) MC results for the case under $-3\%$ strain: the spin structure factor $S(\textbf{k})$ for $A^-$ order and specific heat $C_{\rm v}$ as a function of $T$. (d) The real space spin pattern (MC snapshot) at low-$T$.}
\label{Fig4}
\end{figure}

Indeed, a magnetic transition from $A^+$ and $A^-$ can be realized by applying biaxial strain to LiFe$_2$F$_6$. Fig.~\ref{Fig4} (a) shows the energy differences as a function of strain (defined as $\delta=\frac{a'}{a}-1$ where $a'$ is the strained in-plane lattice constant). Once the compressive strain is beyond $-0.5\%$, the $A^-$ state eventually has the lowest energy, and hence becomes the ground state. By fitting the DFT energy using the Heisenberg model, the coefficients $J_1$ and $J_2$ are extracted as a function of strain, as shown in Fig.~\ref{Fig4}(b). It is clearly that the sign of $J_1$ is changed by strain, while $J_2$ is very robust. The sign change of $J_1$ is associated with the change of bond angle of $J_1$ path (Fe$^{2+}$-F-Fe$^{3+}$), which increases from $97.5^{\circ}$ ($\delta=3\%$) to $102.2^{\circ}$ ($\delta$=$-3\%$), as shown in the insert of Fig.~\ref{Fig4}(b). For the $\delta$=$-3\%$ case, the single-axis magnetocrystalline anisotropy remains along the $c$-axis ($A$=$0.254$ meV/Fe) and the exchange interactions become: $J_1$=$-2.6$ meV, $J_2$=$-11.3$ meV, which lead to $T_{\rm N}$=$110$ K for the $A^-$ state (Fig.~\ref{Fig4}(c-d)). Although this ferrimagnetism is still below room-$T$, it is beyond the boiling point of liquid nitrogen and much higher than the ferromagnetic $T_{\rm C}$ ($4$ K) in strained EuTiO$_3$ \cite{Lee:Nat}.

\begin{figure}
\centering
\includegraphics[width=0.45\textwidth]{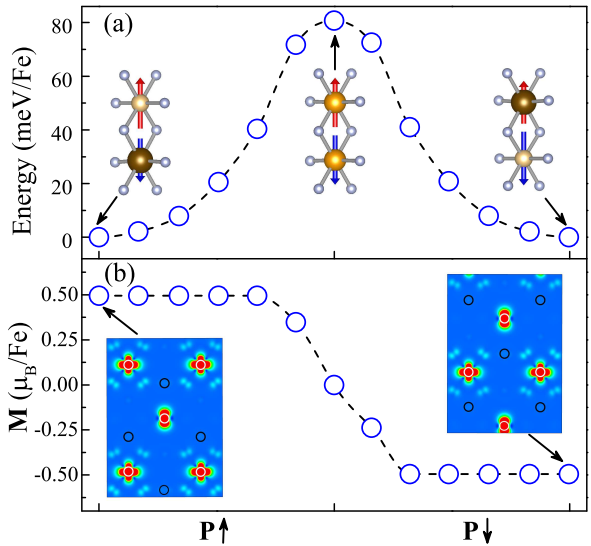}
\caption{Sketch of CO mediated magnetoelectricty in strained ($-3\%$) LiFe$_2$F$_6$. (a) Switch of ferroelectric $\textbf{P}$ simulated by the NEB method. The simulated energy barrier for switching should be considered as the upper-limit in real experiment, while other paths/processes with lower barriers are possible. Inserts: the initial, intermediate, and final structures.(b) The corresponding switch of magnetic $\textbf{M}$ obtained in the NEB process. Inserts: the corresponding profiles (viewed from the [110] direction.) of partial charge density for the topmost valence band. White/black circle: Fe$^{2+}$/Fe$^{3+}$.}
\label{Fig5}
\end{figure}

Since both $\textbf{P}$ and $\textbf{M}$ are in proportional to the CO parameter $C$, this common origin guarantees the intrinsically strong magnetoelectricity and equivalent level of $\textbf{M}$ and $\textbf{P}$. If the antiferromagnetic $\textbf{L}$ is conserved, the tuning of $C$ can change $\textbf{P}$ and $\textbf{M}$ simultaneously. Based on this mechanism, the ferrimagnetic $\textbf{M}$ can be switched up/down accompanying $\textbf{P}$ via electric method, giving the desired electric-control-magnetism function expressed as $(C\cdot\textbf{c})(\textbf{M}\cdot\textbf{L})$, where $\textbf{c}$ is the vector along the polarization direction.

In traditional ferroelectrics, e.g. BaTiO$_3$, the switching of $\textbf{P}$ is by simply moving the ions. However, in electronic ferroelectrics like LiFe$_2$F$_6$, the process of ferroelectric switching may be different. A possible route is that the extra electron of Fe$^{2+}$ directly hops to Fe$^{3+}$ via quantum tunnelling (electron first), which should be quite fast microscopically. The magnetic order is kept in this tunneling process and then the structure (e.g. F octahedra) gradually relaxes according to the new CO pattern. Both $\textbf{P}$ and $\textbf{M}$ can be switched from positive to negative.

Another possible process is to change the structure first, like what happens in the proper ferroelectrics. Starting from the $+P$ LSS, the material climbs an energy barrier to the HSS ($P=0$) first, then to the $-P$ LSS finally. The electron of Fe$^{2+}$ moves smoothly accompanying the structural change. This process can be simulated using the NEB method, as done in Fig.~\ref{Fig5}.

If the real material adopts the second route (structure first), there is another uncertainty regarding the magnetism. For the intermediate HSS, the $A^+$ state is always lower in energy than $A^-$ due to the magnetostriction, even under the compressive condition. Therefore, it is uncertain the switching process will become LSS+$A^-$ to HSS+$A^+$ to LSS+$A^-$ (this process can not be simulated by the NEB method). If so, the antiferromagnetic $A^-$ order is broken in the middle process, and thus the flipping of $M$ becomes uncertain. However, since the switching in such an electronic ferroelectric material is a kinetic process rather than equilibrium process, it is questionable whether the HSS+$A^+$ can really happen since it needs a relative long time for magnetostrictive relaxation (see Supplemental Materials for more discussions). In the current stage, our theoretical methods can not simulate the dynamics of magnetoelectric switching in LiFe$_2$F$_6$. Thus further experiments are encouraged to verify the real process.

Another ferroelectric ferrimagnetic material, Zn$_2$FeOsO$_6$, was predicted recently  \cite{PSWang:Prl}. However, it is a proper ferroelectric material and thus conceptually different from LiFe$_2$F$_6$. From the viewpoint of potential applications, the element Os is very expensive and toxic. And Zn$_2$FeOsO$_6$ is still to be synthesized.

\textit{Conclusion.-} The physical properties of LiFe$_2$F$_6$ were theoretically investigated. The $A^+$-type antiferromagnetism was confirmed to be the ground state. The charge ordering of Fe$^{2+}$/Fe$^{3+}$ configuration can lead to room-temperature ferroelectricity. More interestingly, the ferrimagnetic $A^-$ state with a net magnetization $0.5$ $\mu_{\rm B}$/Fe can be stabilized by moderate compressive strain. In this single phase multiferroic system with both large polarization and magnetization, the intrinsic and strong magnetoelectric coupling can be mediated by charge ordering. It is expected to flip the net magnetization together with the polarization by electric voltage, which provides the desired magnetoelectric function in practice. Further study on other Li$M_2$F$_6$ ($M$ is a $+2$/$+3$ transition metal ion) systems is encouraged to pursuit more multiferroics with better performance.

\acknowledgments{The authors are grateful to Y. K. Tang, X. H. Niu, Y. H. Li, and L. Shi for illuminating discussions. This work was supported National Natural Science Foundation of China (Grant No. 11674055), the Fundamental Research Funds for the Central Universities, and Jiangsu Innovation Projects for Graduate Student (Grant No. KYCX17\_0047). Most calculations were supported by National Supercomputer Center in Guangzhou (NSCC-GZ).}

\bibliographystyle{apsrev4-1}
\bibliography{ref3}

\newpage
\section{Supplemental Materials}
\subsection{Test of $U_{\rm eff}$ for LiFe$_2$F$_6$}
\begin{figure}
\centering
\includegraphics[width=0.48\textwidth]{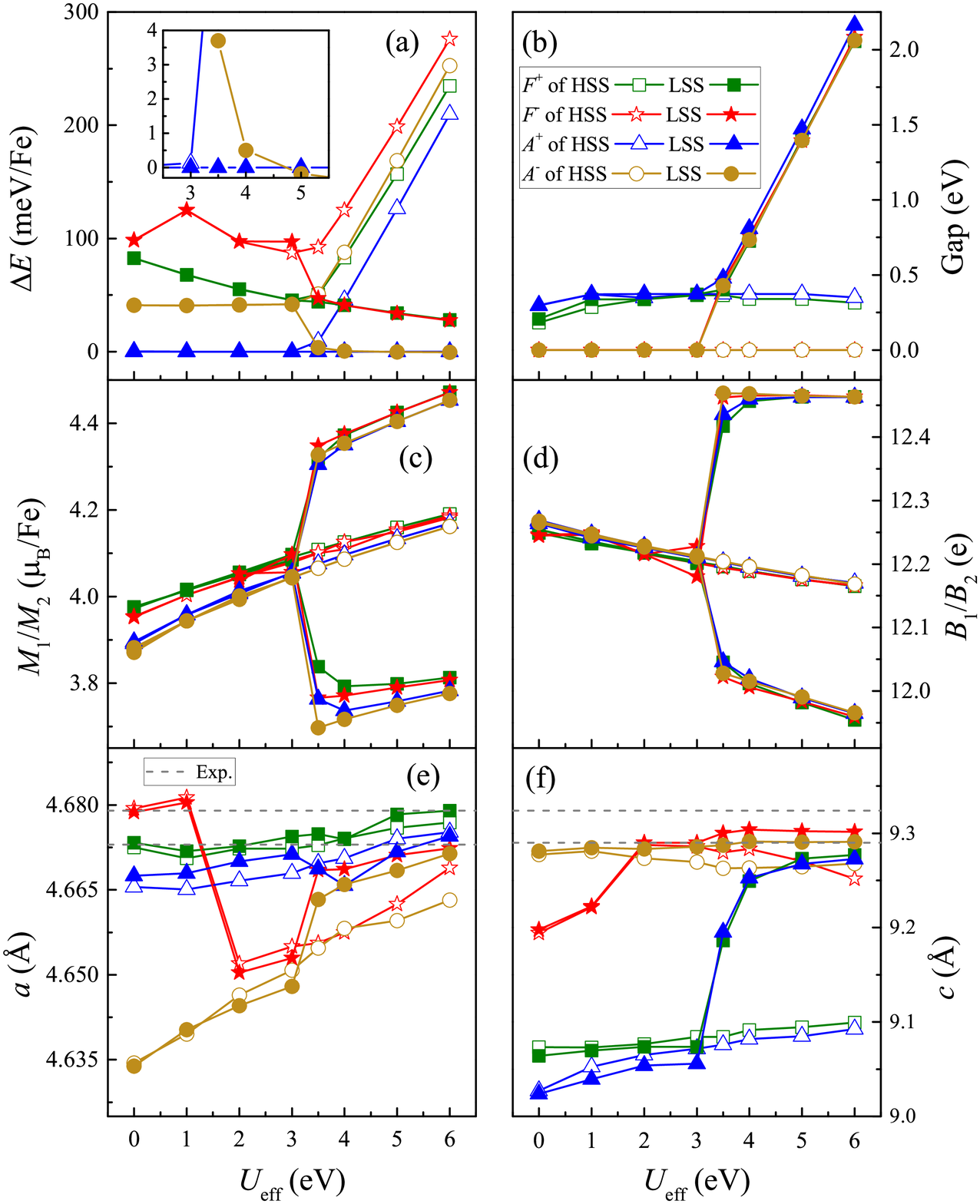}
\caption{Calculated physical properties for different magnetic orders as a function of $U_{\rm eff}$. (a) Energy differences ($\Delta E$'s). The $A^+$ of LSS is taken as the reference of energy. (b) Band gaps. (c) Local magnetic moment ($M_1$ for Fe$1$ and $M_2$ for Fe$2$) within the default PAW sphere. (d) Bader charge ($B_1$ for Fe$1$ and $B_2$ for Fe$2$). (e-f) Lattice constants $a$/$c$ in the tetragonal notation. The experimental values (Exp. for short) of lattice constants are also shown in (e-f) for comparison. Both the HSS and LSS are calculated.}
\label{FigS1}
\end{figure}

To obtain a suitable $U_{\rm eff}$ value, we have calculated several physical properties as a function of $U_{\rm eff}$. As shown in Fig.~\ref{FigS1}, the CO occurs when $U_{\rm eff}\geqslant3.5$eV and the $A^+$ antiferromagnetic is the ground magnetic state only when $U_{\rm eff}\leqslant4$eV. Considering the existence of Fe$^{2+}$ and Fe$^{3+}$ in LiFe$_2$F$_6$ found by M\"{o}ssbauer spectrum measurement \cite{Greenwood:Jcsa,Fourquet:Jssc} and the $A^+$ magnetic ground state detected by neutron powder diffraction \cite{Shachar:Prb,Wintenberger:Ssc}, $3.5\leqslant U_{\rm eff}\leqslant4$ eV could be a suitable range. In addition, the lattice constants obtained with $U_{\rm eff}$ in this range are close to the experimental one.

Although $U_{\rm eff}=3.5$ eV can also lead to $A^+$ magnetic ground state and CO, the N\'eel temperature estimated using the exchanges at this $U_{\rm eff}=3.5$ eV deviates from the experimental one more than that at $U_{\rm eff}=4$ eV. Thus $U_{\rm eff}=4$ eV could be the most suitable value, which is used as default in our calculations.

In fact, in our previous DFT calculations, $U_{\rm eff}=4$ eV was tested to be the best choice for Fe$^{2+}$/Fe$^{3+}$-based perovskites.

\begin{figure}
\centering
\includegraphics[width=0.48\textwidth]{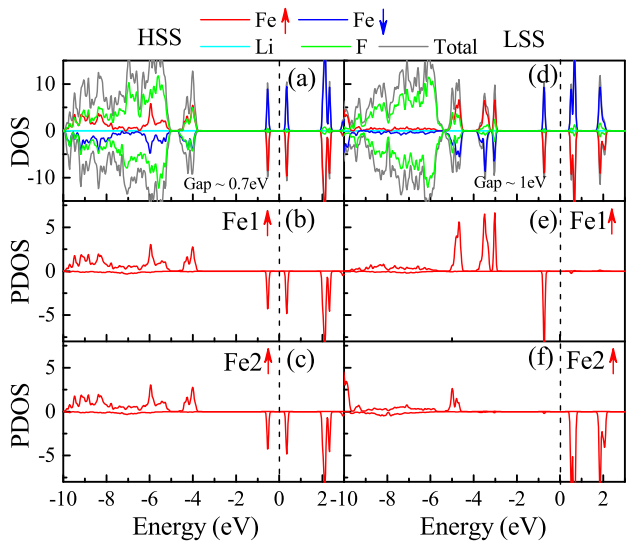}
\caption{DOS and PDOS of $A^+$ calculated by the hybrid functional based on HSE06 exchange. (a-c) For the $P4_2/mnm$ HSS. (d-f) For the $P4_2nm$ LSS. $\uparrow$/$\downarrow$ denote the spin directions.}
\label{FigS2}
\end{figure}

Fig.~\ref{FigS2} shows the DOS and PDOS of unstrained LiFe$_2$F$_6$, obtained in the HSE06 hybrid functional calculation.

\subsection{The $A^+$ state of the high symmetric structure}
According to data shown in Table I of main text, for the LSS, the energies of the $A^+$ and $A^-$ (also $F^+$ and $F^-$) states are proximate, implying the very weak $J_1$ exchange. However, for the high symmetric case, this energy difference becomes quite large ($\sim41$ meV/Fe). And the lattice constant along the $c$ axis is abnormally short for the $A^+$ and $F^+$ cases, comparing with other calculated ones and experimental ones. Furthermore, the band gap is significantly different between $A^+$ and $A^-$ (also $F^+$ and $F^-$). These differences suggest nontrivial physical mechanism hidden in the high symmetric cases.

\begin{figure}
\centering
\includegraphics[width=0.48\textwidth]{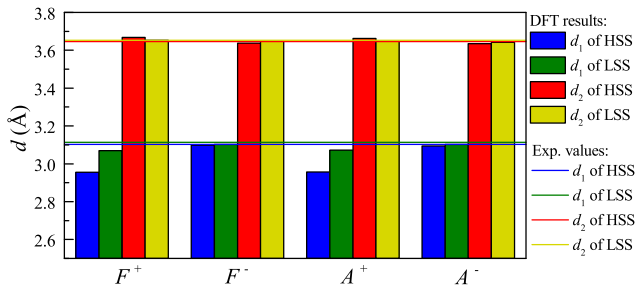}
\caption{The Fe-Fe distances of $J_1$/$J_2$ path ($d_1$/$d_2$) for different magnetic configurations in the optimized HSS and LSS. The experimental values are also shown as lines.}
\label{FigS3}
\end{figure}

By analysing the optimized structures (Fig.~\ref{FigS3}), it is found that the Fe-Fe distances of $J_2$ path are almost identical for all magnetic cases (the difference is within $0.01$ {\AA}), which are very close to the experimental values. But the Fe-Fe distances of $J_1$ path is significantly shorten ($\sim 0.15$ {\AA} shorter than others and experimental values) in the $A^+$ and $F^+$ case of HSS. This structural Fe-Fe dimerization is associated with the magnetostriction. In the high symmetric case without CO, for all Fe's, the valence is uniform $+2.5$. The half integer electron per Fe, can lead to the Peierls transition, which can open a band gap and lower the electronic energy. This is just what happens in the $A^+$ and $F^+$ case of HSS.

Then one will ask why is the Peierls transition absent in the $A^-$ and $F^-$ cases of HSS. In these two cases, the antiferromagnetic coupling between nearest-neighbor Fe's forbids the double-exchange electron hopping, making the half-filling band at the Fermi level extremely narrow. Thus the electronic energy gain from band-opening is negligible. Therefore, the structural dimerization between Fe$\uparrow$-Fe$\downarrow$ is suppressed.

This phenomenon, i.e. large structural change between $A^+$ and $A^-$ (also $F^+$ and $F^-$), is a typical magnetostrictive effect. In the LSS case, the pair of Fe$^{2+}$ and Fe$^{3+}$ already breaks the half-filling condition for dimerization. Thus, this Peierls transition induced  magnetostrictive effect does not exist either in the LSS.

\subsection{Estimation of $J_{\rm CO1}$ \& $J_{\rm CO2}$}
The understanding of this magnetostriction is important for the estimation of CO temperature and magnetoelectricity, as discussed in the following.

For real material of LiFe$_2$F$_6$, the HSS can only exist above $T_{\rm CO}$, in which there is not any magnetic order. Therefore, the magnetostrictive energy gained in the $A^+$ (or $F^+$) calculation of HSS does not exist in real material.

Strickly speaking, the paramagnetic CO and CU state should be compared to calculate $J_{\rm CO1}$ and $J_{\rm CO2}$. However, in principle, the DFT technique can only handle the zero-$T$ situation. It is well known that the non-magnetic DFT calculation can not simulate the paramagnetic materials with disordered local moments, like LiFe$_2$F$_6$ here. As an alternative, here the $A^-$ (or $F^-$) state is adopted to calculate $J_{\rm CO1}$ and $J_{\rm CO2}$, in which the magnetostrictive contribution is negligible. The energy gain for the CO transition is $85.5\pm1.5$ meV/Fe, which is consistent between the $A^-$ and $F^-$.

We also tried other CO state. However, none could be stabilized in the self-consistent DFT calculation, except the experimental one. Thus, only the coefficient ($2J_{\rm CO2}+\frac{J_{\rm CO1}}{2}$) of the charge lattice model can be extracted, while the individual values of $J_{\rm CO2}$ and $J_{\rm CO1}$ are not available.

It is reasonable to argue that the energy gain of CO is mostly from the Coulombic potential between positive and negative charges. The Coulombic energy should be inversely proportional to the distance $d$ between charges, namely $\sim(\frac{1}{d})^n$. For naked charges, $n=1$. In solids, due to complex ionic environment and charge screening, $n$ can be larger. In this sense, $J_{\rm CO1}$ should be larger than $J_{\rm CO2}$, considering $d_2=1.174d_1$. Then in our MC calculation (Fig. 3 in main text), the ratio of $\frac{J_{\rm CO1}}{J_{\rm CO2}}$ are considered from $1$ to $3$. Although the obtained $T_{\rm CO}$ varies as a function of $\frac{J_{\rm CO1}}{J_{\rm CO2}}$, it is above room-temperature in the considered range, in agreement with experimental results. Further experiments are needed to measure the exact $T_{\rm CO}$. Then the effective $\frac{J_{\rm CO1}}{J_{\rm CO2}}$ can be fixed.

\subsection{Estimation of $J_{1}$ \& $J_{2}$}
The magnetostrictive effect always exists in the DFT calculation, more or less. To exclude this contribution and to extract the pure exchange interaction, the exchange interactions are extracted by comparing the energies of different magnetic orders with an identical structure. The optimized structure for the ground magnetic configuration is used. There are four magnetic configurations ($A^+$, $A^-$, $F^+$, and $F^-$) but three are enough to extract $J$'s. We chose the lowest three, but others also gave very close values for $J$'s. Thus, these exchange interactions are not sensitive to the spin configurations and the material can be well described by the Heisenberg Hamiltonian. Physically, the system is a protypical Mott insulator (not a metal), which is an ideal platform for the Heisenberg model.

\subsection{Estimation of magnetocrystalline anisotropy coefficient $A$}
The magnetocrsystalline anisotropy energy (MAE) was calculated by incorporating the spin-orbit coupling (SOC). Such a weak SOC has negligible effects to other quantities, e.g. $J_{1}$/$J_{2}$/$J_{\rm CO}$/energy. Thus, in other calculations, the SOC was not included, although some tests with SOC were done. The MAE is extracted from fully self-consistent calculations of total energy ($E_{a}$, $E_{b}$, and $E_{c}$) for three magnetized directions ($a$, $b$, and $c$-axes) of the ground magnetic state, i.e. $A^+$  for the untrained case and $A^-$  for the strained one. Due to the tetragonal symmetry, $E_{a}$ always equals to $E_{b}$, while $E_{c}$ is the lowest one. So the magnetocrystalline easy axis is along the $c$-axis and the coefficient $A$ can be obtained.

\end{document}